\documentstyle[aps,prl,preprint,floats,epsfig]{revtex}  
\draft 

\def\Journal#1#2#3#4{{#1} {\bf #2}, #3 (#4)}

\def\NPB{{\em Nucl. Phys.} B}
\def\PLB{{\em Phys. Lett.}  B}
\def\PRL{\em Phys. Rev. Lett.}
\def\PRD{{\em Phys. Rev.} D}
\def\ZPC{{\em Z. Phys.} C}

\begin{document}


\preprint{\tighten\vbox{\hbox{\hfil BELLE-CONF-0110}
}}

\title{\large
       Evidence for the Electroweak Penguin Decay
 $B \to X_{s} \ell^{+} \ell^{-}$}

\author{The Belle Collaboration}
\maketitle

\tighten

\begin{abstract}


We report a search for the flavor-changing neutral current decay $B \to X_{s} \ell^{+} \ell^{-}$ using a 29.5~fb${}^{-1}$ data sample accumulated at the $\Upsilon(4S)$ resonance with Belle detector at the KEKB $e^{+}e^{-}$ storage ring. We observe evidence for $B\to K\mu^+\mu^-$ and report the preliminary branching fraction of 
\begin{center}
	${\cal B}(B \to K \mu^{+} \mu^{-}) = (0.99^{+0.39}_{-0.32}{}^{+0.13}_{-0.15}) \times 10^{-6}$.

\end{center}
We also set the following 90\% confidence level upper limits on the branching fractions for exclusive and inclusive decays,
\begin{center}
	${\cal B}(B \to K e^{+} e^{-}) < 1.2 \times 10^{-6}$,\\
	${\cal B}(B \to K^*(892) e^{+} e^{-}) < 5.1 \times 10^{-6}$,\\
	${\cal B}(B \to K^*(892) \mu^{+} \mu^{-}) < 3.0 \times 10^{-6}$,\\
	${\cal B}(B \to X_{s} e^{+} e^{-}) < 10.1 \times 10^{-6}$,\\
	${\cal B}(B \to X_{s} \mu^{+} \mu^{-}) < 19.1 \times 10^{-6}$.
\end{center}

\end{abstract}
  

\newpage

\begin{center}
  K.~Abe$^{9}$,               
  K.~Abe$^{37}$,              
  R.~Abe$^{27}$,              
  I.~Adachi$^{9}$,            
  Byoung~Sup~Ahn$^{15}$,      
  H.~Aihara$^{39}$,           
  M.~Akatsu$^{20}$,           
  K.~Asai$^{21}$,             
  M.~Asai$^{10}$,             
  Y.~Asano$^{44}$
% ...
\end{center}

\small
\begin{center}
% ...
$^{9}${High Energy Accelerator Research Organization (KEK), Tsukuba}\\
$^{10}${Hiroshima Institute of Technology, Hiroshima}\\
% ...
$^{15}${Korea University, Seoul}\\
% ...
$^{20}${Nagoya University, Nagoya}\\
$^{21}${Nara Women's University, Nara}\\
% ...
$^{27}${Niigata University, Niigata}\\
% ...
$^{37}${Tohoku Gakuin University, Tagajo}\\
% ...
$^{39}${University of Tokyo, Tokyo}\\
% ...
$^{44}${University of Tsukuba, Tsukuba}\\
% ...
\end{center}

\normalsize



\section{Introduction}
Flavor-changing neutral current (FCNC) processes are forbidden at the tree level in the Standard Model (SM), but are induced by loop or box diagrams.
If non-SM particles participate in the loop or box diagrams,
their amplitudes may interfere with the SM amplitudes. This makes FCNC
processes an ideal place to search for new physics.

CLEO first observed and measured the radiative penguin decay $B\to X_{s}\gamma$\cite{CLEO-radiative}, which constrains the magnitude of the effective Wilson coefficient of the electromagnetic penguin operator, $|C_7^{\mathrm{eff}}|$.
This provides the most stringent indirect limit on the charged Higgs mass range\cite{charged-Higgs}.
However, it cannot constrain the phase of $C_7^{\mathrm{eff}}$, which is essential to obtain definitive evidence of new physics since $C_7^{\mathrm{eff}}$ is positive in the SM while it can be negative in non-SM physics\cite{C7}.
In particular, $C_7^{\mathrm{eff}}$ is constrained to be negative in the minimal SUGRA model\cite{MSUGRA}.
The electroweak penguin decay $B \to X_{s} \ell^{+} \ell^{-}$ is promising from this point of view since the coefficients $C_7^{\mathrm{eff}},\ C_9^{\mathrm{eff}}$ and $C_{10}$ can be completely determined by measuring the dilepton invariant mass distributions, forward-backward charge asymmetry of the dilepton together with the $B\to X_{s}\gamma$ decay rate\cite{Ali-1994}.

The branching fractions predicted within the framework of the Standard Model are listed in Tables~\ref{tab:brpred} and \ref{tab:brpredinc} [\ref{ref:Ali}--\ref{ref:inclusive-KS}].
Although several groups \cite{Babar}\cite{CDF}\cite{CLEO} have searched for exclusive $B\to K\ell^+\ell^-$ and $B\to K^*(892)\ell^+\ell^-$ decays\footnote{$K^*(892)$ is referred as $K^*$ hereafter.} and CLEO\cite{CLEO-inclusive} has searched for inclusive $B \to X_{s} \ell^{+} \ell^{-}$ decays,
no evidence has been observed. 

\begin{table}[htpb]
\caption{Branching fractions for $B\to K\ell^+\ell^-$ and $B\to K^{*}\ell^+\ell^-$ decays predicted in the Standard Model.}
\begin{center}
\begin{tabular}{llll}
& \multicolumn{3}{c}{Predicted branching fraction $[\times 10^{-6}]$} \\ \cline{2-4}
\raisebox{1.5ex}[0pt]{Mode} & Ali {\it et al.}\cite{Ali}   & Greub {\it et al.}\cite{Greub}& Melikhov {\it et al.}\cite{Melikhov} \\ \hline
$K^{*} e^{+} e^{-} $      & $2.3^{ + 0.7}_{ - 0.4}$ & $1.4 \pm 0.3$  & $1.4 \pm 0.5$ \\ 
$K^{*} \mu^{+} \mu^{-} $  & $1.9^{ + 0.5}_{ - 0.3}$ & $1.0 \pm 0.2$  & $1.0 \pm 0.4$ \\  
$K e^{+} e^{-} $          & $0.57^{ + 0.16}_{ - 0.10}$ & $0.33 \pm 0.07$  & $0.42 \pm 0.09$ \\ 
$K \mu^{+} \mu^{-} $      & $0.57^{ + 0.16}_{ - 0.10}$ & $0.33 \pm 0.07$  & $0.42 \pm 0.09$ \\ 
\end{tabular}
\end{center}
\label{tab:brpred}
\end{table}

\begin{table}[htpb]
\caption{Branching fractions for inclusive $B\to X_s\ell^+\ell^-$ decays predicted in the Standard Model.}
\begin{center}
\begin{tabular}{lll}
& \multicolumn{2}{c}{Predicted branching fraction $[\times 10^{-6}]$} \\ \cline{2-3}
\raisebox{1.5ex}[0pt]{Mode} & Ali {\it et al.}\cite{inclusive-BF-Ali} & Kr\"uger {\it et al.}\cite{inclusive-KS}\\ \hline
$X_{s} e^{+} e^{-}$   	& $8.4\pm2.3$ & N/A \\
$X_{s} \mu^{+} \mu^{-}$	& $5.7\pm1.2$ & 6.7 \\
\end{tabular}
\end{center}
\label{tab:brpredinc}
\end{table}

In this paper, we present the preliminary results of a search for $B$ decays to an oppositely charged lepton pair and a strange hadron system using data produced in $e^+{}e^{-}$ annihilation at the KEKB asymmetric collider \cite{kekb}, and collected with the Belle detector. The data sample corresponds to 29.5~fb${}^{-1}$ taken at the $\Upsilon(4S)$ resonance and contains approximately 31.3 million $B\overline{B}$ pairs.

Belle is a general-purpose detector based on a 1.5~T superconducting solenoid magnet that surrounds the KEKB beam crossing point. Charged particle tracking covering approximately 90\% of the total cm solid angle is provided by  a Silicon Vertex Detector (SVD), consisting of three nearly cylindrical layers of double-sided silicon strip detectors~\cite{SVD}, and a 50-layer Central Drift Chamber (CDC)\cite{CDC}. 
Impact parameter resolutions are measured as functions of momentum $p$ (GeV/$c$) to be $\sigma_{xy}=19 \oplus 50/(p\beta\sin^{3/2}\theta)$~$\mu$m and $\sigma_{z}=36 \oplus 42/(p\beta\sin^{5/2}\theta)$~$\mu$m, where $\theta$ is the polar angle with respect to the beam direction.
The transverse momentum resolution for charged tracks is
$(\sigma_{p_T}/p_T)^2=(0.0019 p_T)^2+(0.0030/\beta)^2$, where $p_T$ is in GeV/$c$.
Particle identification is accomplished by a combination of a silica Aerogel Cherenkov Counters (ACC)\cite{ACC}, a Time of Flight counter system (TOF)\cite{TOF} and $dE/dx$ measurements in the CDC. 
The combined response of the three systems provide $K^\pm$ identification with an efficiency of about 85\% and a charged pion fake rate of about 10\% for all momenta up to 3.5~GeV/$c$.
A CsI(Tl) Electromagnetic Calorimeter (ECL) located inside the solenoid coil is used for $\gamma/\pi^0$ detection and electron identification\cite{ECL}.
The photon energy resolution is $(\sigma_E/E)^2=(0.013)^2+(0.0007/E)^2+(0.008/E^{1/4})^2$, where $E$ is in GeV.
The $\mu/K_L$ detector (KLM)\cite{KLM} is located outside of the coil. An Extreme Forward Calorimeter (EFC)\cite{EFC} is situated near the beam pipe.
A detailed description of the Belle detector can be found elsewhere\cite{NIM}.

\section{Signal Property}

To estimate the signal detection efficiency, we generate exclusive and inclusive simulated signal events. The exclusive $B\to K\ell^+\ell^-$ and $B\to K^*\ell^+\ell^-$ Monte Carlo (MC) sample is generated according to a model described by Greub, Ioannissian and Wyler\cite{Greub}. 
The $K^{*} \ell^+\ell^-$ decay has a pole at $q^2 = 0$ since a nearly real photon couples to dileptons, while $K\ell^+\ell^-$ does not have a pole at $q^2 = 0$ due to helicity suppression.

The inclusive $B\to X_s\ell^+\ell^-$ MC sample is generated according to an $s$-distribution with the lepton mass term modeled by Kr\"uger and Sehgal\cite{inclusive-KS}, and a $u$-distribution modeled by Ali {\it et al.}\cite{AliInc}.
The exclusive MC sample described above is used to account for resonant states in the region $M_{X_{s}}<1.0$~GeV/$c^2$.
The fractions for exclusive decays are determined from the inclusive branching fractions predicted by Ali, Hiller, Handoko and Morozumi\cite{inclusive-BF-Ali} and the exclusive branching fractions predicted by Ali, Ball, Handoko and Hiller\cite{Ali} shown in Tables~\ref{tab:brpred} and \ref{tab:brpredinc}. The interference between $B\to X_s\ell^+\ell^-$ and long distance processes $B\to J/\psi(\psi^{'}) X_s$ is not considered for either MC sample.

\section{Analysis}
\subsection{Exclusive Analysis}
In $B \to K^{(*)} \ell^{+} \ell^{-} $ decays, the K(*) hadronic system and the two oppositely charged leptons form a quasi-three body state.
In the $K \ell^{+} \ell^{-}$ mode, the hadronic system is one kaon, while in the $K^{*} \ell^{+} \ell^{-}$ mode, the hadronic system contains one kaon and a pion. 

The distance of the closest approach to the interaction point of the charged track is required to be less than 0.5~cm in the $r\phi$ plane and less than 5.0~cm in the $z$ direction.
This requirement reduces the combinatorial background from photon conversion, beam-gas and beam-wall events. 
Electrons are identified from the ratio of shower energy in the ECL to the momentum measured by the CDC, the shower shape of the cluster in the ECL, the energy deposit in the CDC and the light yield in ACC.
Tracks are identified as muons by their penetration length in the KLM and the matching between the tracks found by the CDC and hits in the KLM.
To reduce the misidentification of hadrons as leptons, we require that the momentum be greater than $0.5$~GeV/$c$ and $1.0$~GeV/$c$ for electron and muon candidates, respectively.

A charged $K/\pi$ is identified by a likelihood ratio based on the $dE/dx$ in the CDC, time-of-flight information and ACC response.

Photons are selected from isolated showers in the ECL whose energy is greater than $50$~MeV and whose shape is consistent with electromagnetic shower. Neutral pion candidates are reconstructed from pairs of these photons and required to have the invariant mass within $10$~MeV/$c^2$ of the nominal $\pi^{0}$ mass. $K^{0}_{S}$ candidates are reconstructed from oppositely charged tracks whose vertex position is displaced from the interaction point. We require the invariant mass lies within $15$~MeV/$c^{2}$ of the nominal $K^{0}_{S}$ mass.

$K^{*}$ candidates are formed by combination of a charged or neutral kaon and a pion ($K^{+}\pi^{-}$, $K^{0}_{S}\pi^{0}$, $K^{0}_{S}\pi^{+}$ or $K^{+}\pi^{0}$).\footnote{Charge conjugate modes are implied throughout this paper.} The $K^{*}$ invariant mass is required to lie within $75$~MeV/$c^2$ of the nominal $K^{*}$ mass. This value corresponds to 1.5 times the $K^{*}$ natural width. 
For modes involving $\pi^{0}$'s, there are large combinatorial backgrounds due to the abundance of low momentum $\pi^{0}$'s. To reduce such backgrounds, we require that the $K^{*}$ helicity angle $\cos\theta_{\rm{hel}}$, defined as the angle between $K^{*}$ momentum direction and kaon momentum direction in the $K^{*}$ rest frame, be less than 0.8.

$B$ candidates are reconstructed from the $K^{(*)}$ hadron system and an oppositely-charged lepton pair. 
The backgrounds from the long distance processes $J/\psi(\psi^{'}) K^{(*)}$ are rejected using the dilepton invariant mass. The veto windows are defined as,
\begin{eqnarray*}
  -0.25 &<& M_{e^{+}e^{-}}     - M_{J/\psi}   < 0.07~\textrm{GeV/}c^{2}    {\hskip 1cm}  \textrm{for $K^{*}$ modes} \\
  -0.20 &<& M_{e^{+}e^{-}}     - M_{J/\psi}   < 0.07~\textrm{GeV/}c^{2}    {\hskip 1cm}  \textrm{for $K$ modes} \\
  -0.20 &<& M_{e^{+}e^{-}}     - M_{\psi^{'}} < 0.07~\textrm{GeV/}c^{2}    {\hskip 1cm}  \textrm{for $K^{*}$ and $K$ modes}\\
  -0.15 &<& M_{\mu^{+}\mu^{-}} - M_{J/\psi}   < 0.08~\textrm{GeV/}c^{2}     {\hskip 1cm}  \textrm{for $K^{*}$ modes}\\
  -0.10 &<& M_{\mu^{+}\mu^{-}} - M_{J/\psi}   < 0.08~\textrm{GeV/}c^{2}     {\hskip 1cm}  \textrm{for $K$ modes}\\
  -0.10 &<& M_{\mu^{+}\mu^{-}} - M_{\psi^{'}} < 0.08~\textrm{GeV/}c^{2}     {\hskip 1cm}  \textrm{for $K^{*}$ and $K$ modes}
\end{eqnarray*}
To suppress background from photon conversions or $\pi^{0}$ Dalitz decays, we require the dielectron mass to satisfy $M_{e^{+}e^{-}} > 0.14~$GeV/$c^{2}$.

The backgrounds from continuum $q\overline{q}$ events are suppressed using event shape variables. Continuum events have a jet-like shape while $B\overline{B}$ events have a spherical shape in the center-of-mass frame.
A Fisher discriminant $\cal{F}$\cite{fd} is calculated from the energy flow in 9 cones along the candidate sphericity axis and the second Fox-Wolfram moment $R_{2}$\cite{fw}. 
In addition to the Fisher discriminant, the $B$ meson flight direction $\cos\theta_{B}$ and the angle between $B$ meson sphericity axis and the $z$ axis, $\cos\theta_{\rm{sph}}$, are used to suppress the continuum events. 
For the muon mode, $|\cos\theta_{\rm{sph}}|$ is not used since its distribution is nearly the same for signal and continuum due to detector acceptance. We combine $\cal{F}$, $\cos\theta_{B}$ and $\cos\theta_{\rm{sph}}$ into one likelihood ratio ${\cal{LR}}_{\rm{cont}}$ defined as,
\begin{eqnarray*}
  {\cal{LR_{\rm{cont}}}} = \frac{\cal{L_{\rm{sig}}}}{\cal{L_{\rm{sig}}} + \cal{L_{\rm{cont}}}}
\end{eqnarray*}
where ${\cal{L_{\rm{sig}}}}$ and ${\cal{L_{\rm{cont}}}}$ are the products of the likelihoods for signal and continuum background, respectively.

Most lepton candidates originate from semileptonic $B$ decays in $B\overline{B}$ events. The missing energy of the event, $E_{\mathrm{miss}}$, is used to suppress the background since we expect large missing energy due to the undetected neutrino.
The $B$ meson flight angle is also used to suppress combinatorial background in $B\overline{B}$ events. 
We combine $E_{\mathrm{miss}}$ and $\cos\theta_{B}$ into the likelihood ratio ${\cal{LR}}_{B\overline{B}}$.

We use the beam constrained mass $M_{bc}$ and the energy difference $\Delta E$ in the center-of-mass frame to select $B$ candidates where
\begin{eqnarray*}
  M_{bc} &=& \sqrt{ s/4 - \left( \sum_{i} p^{*}_{i} \right)^{2}}, \\
  \Delta E &=& \sum_{i} E^{*}_{i} - \sqrt{s} / 2.
\end{eqnarray*}

\begin{figure}
\hspace{5mm}(a)\hspace{3.1 in}(b)\\
\vspace*{-13mm}
\begin{center}
\epsfxsize 2.8 truein \epsfbox{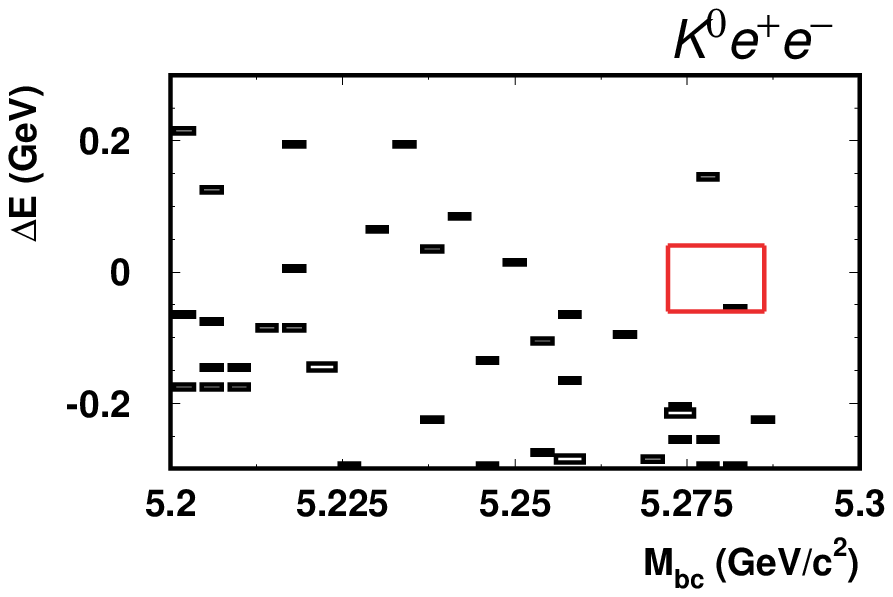}
\hspace{11mm}
\epsfxsize 2.8 truein \epsfbox{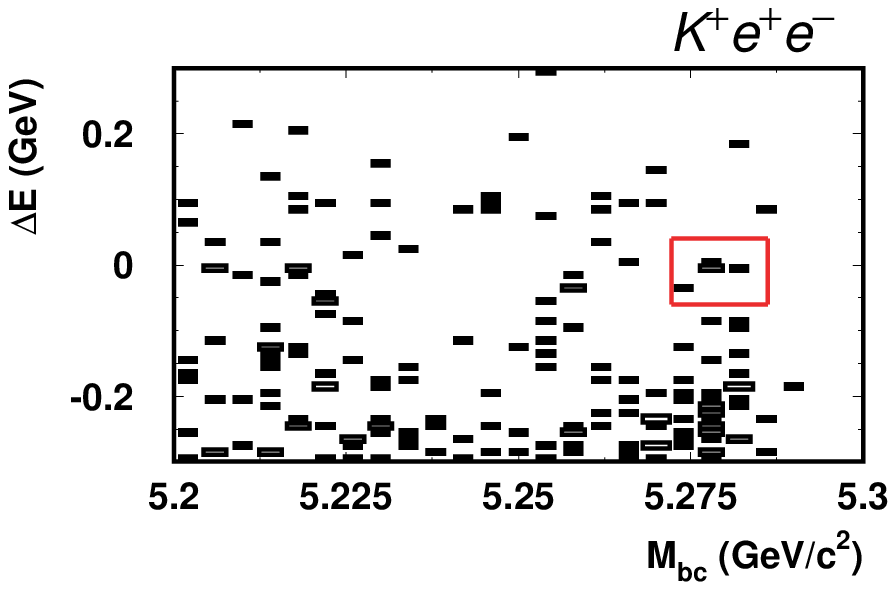}\\
\end{center}
\hspace{5mm}(c)\hspace{3.1 in}(d)\\
\vspace*{-13mm}
\begin{center}
\epsfxsize 2.8 truein \epsfbox{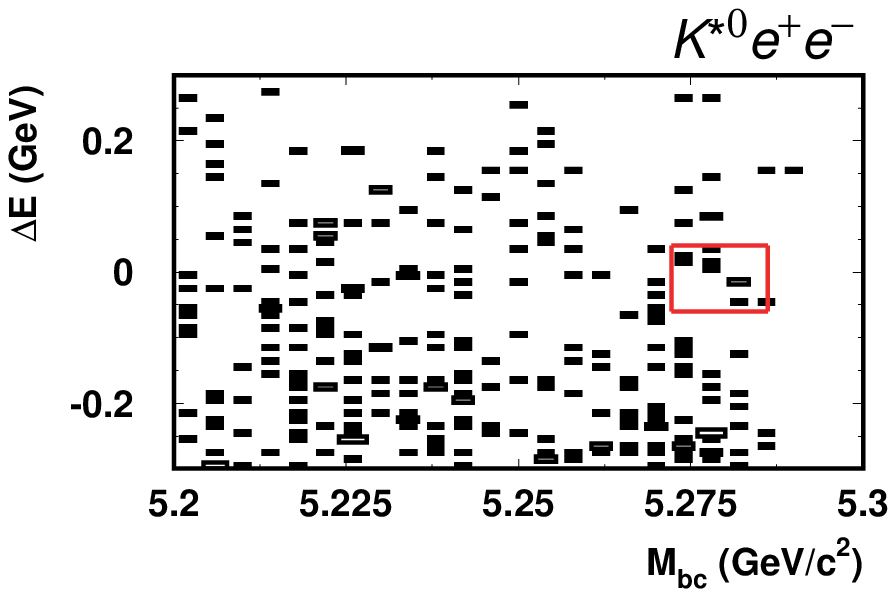}
\hspace{11mm}
\epsfxsize 2.8 truein \epsfbox{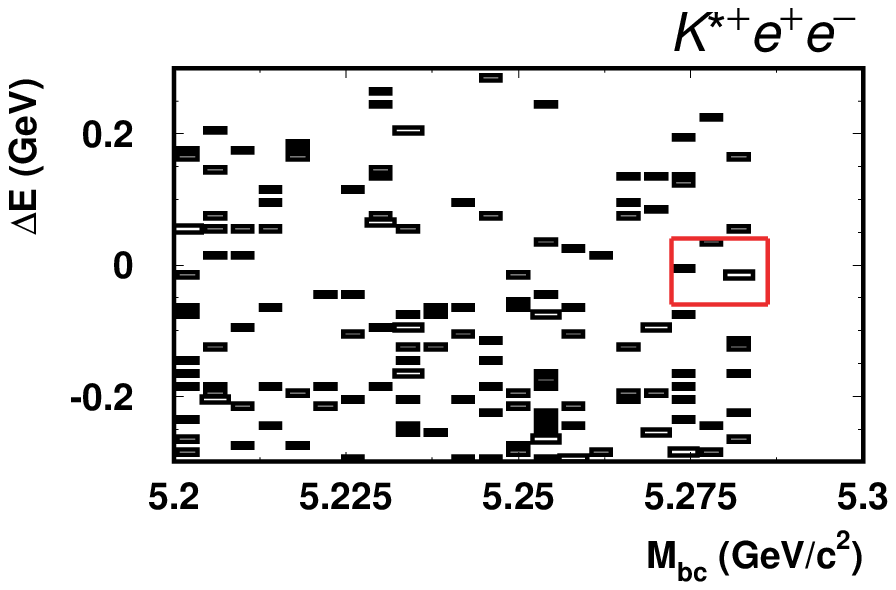}\\
\end{center}
\hspace{5mm}(e)\hspace{3.1 in}(f)\\
\vspace*{-13mm}
\begin{center}
\epsfxsize 2.8 truein \epsfbox{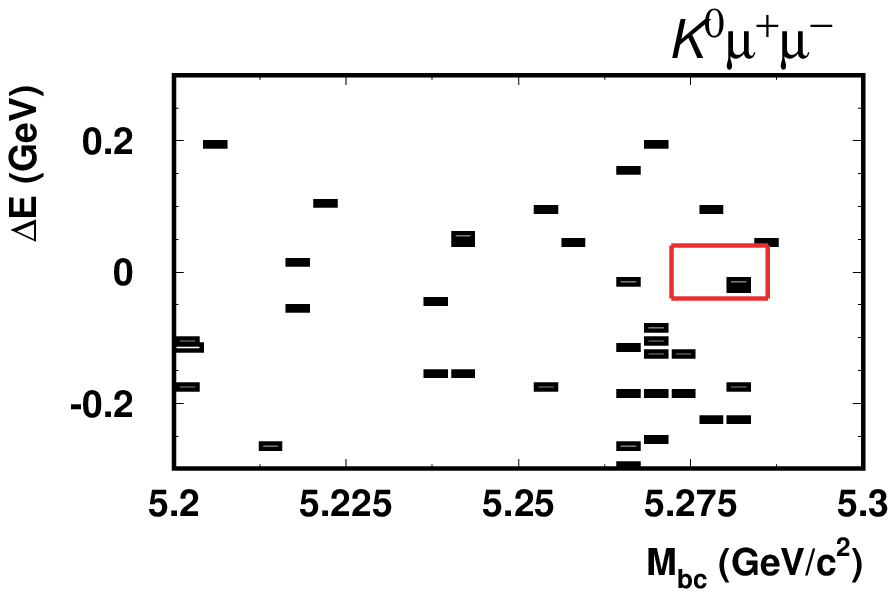}
\hspace{11mm}
\epsfxsize 2.8 truein \epsfbox{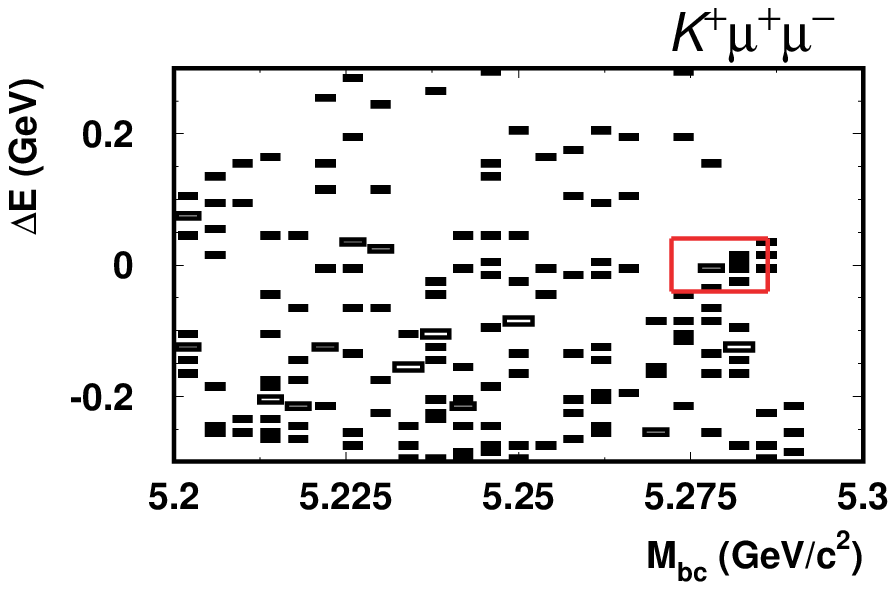}\\
\end{center}
\hspace{5mm}(g)\hspace{3.1 in}(h)\\
\vspace*{-13mm}
\begin{center}
\epsfxsize 2.8 truein \epsfbox{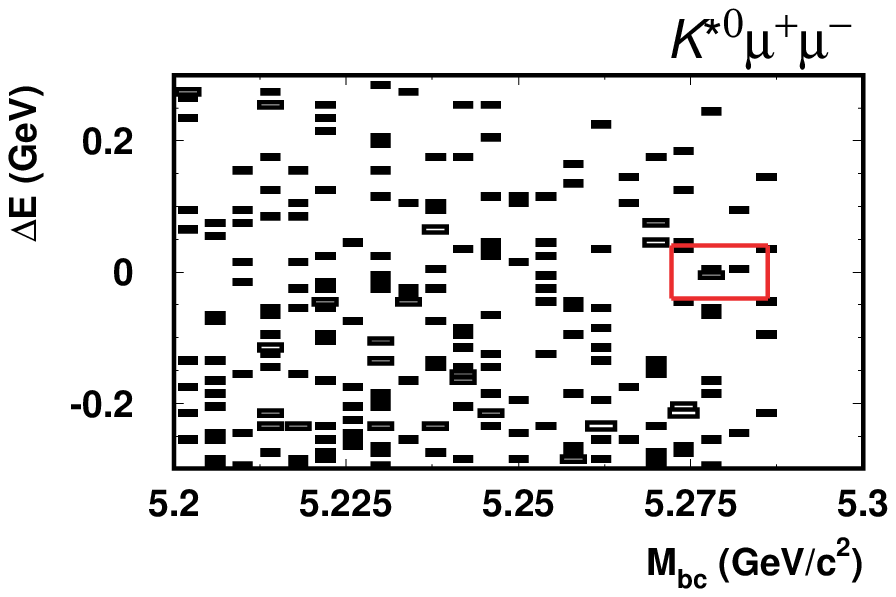}
\hspace{11mm}
\epsfxsize 2.8 truein \epsfbox{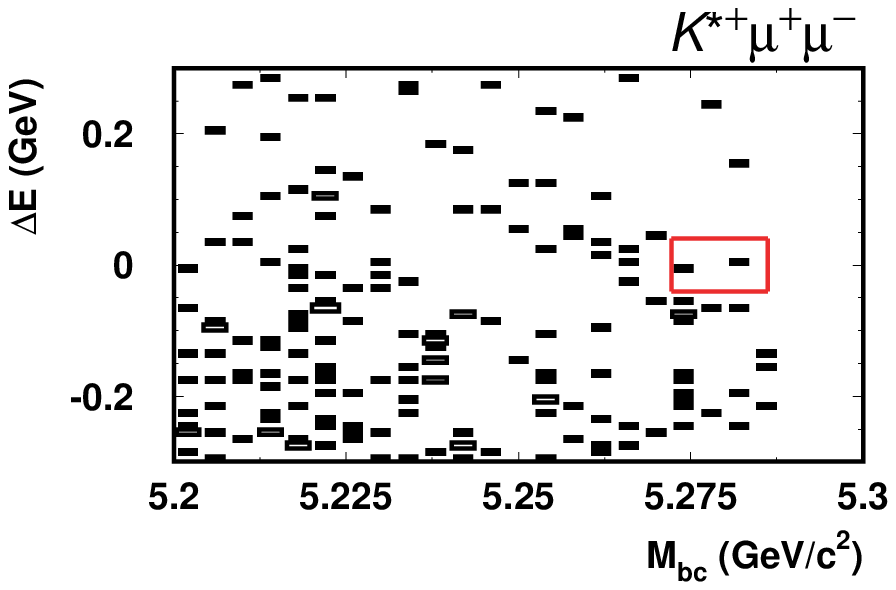}\\
\end{center}
\caption{$\Delta E$ vs. $M_{bc}$ scatter plots for (a) $B^0\to K^{0} e^{+} e^{-}$, (b) $B^+\to K^{+} e^{+} e^{-}$, (c) $B^0\to K^{*0} e^{+} e^{-}$, (d) $B^+\to K^{*+} e^{+} e^{-}$, (e) $B^0\to K^{0} \mu^{+} \mu^{-}$, (f) $B^+\to K^{+} \mu^{+} \mu^{-}$, (g) $B^0\to K^{*0} \mu^{+} \mu^{-}$ and (h) $B^+\to K^{*+} \mu^{+} \mu^{-}$ samples in data.
}
\label{fig:dE-Mb}
\end{figure}

The selection criteria are tuned to maximize the expected significance $S/\sqrt{(S+B)}$ where $S$ is the signal yield and $B$ is the expected background in the signal box. $S$ and $B$ are determined from GEANT based MC samples with an effective luminosity of 40~fb${}^{-1}$, assuming the branching fractions predicted by Ali {\it et al.}\cite{Ali}. The signal box is defined as $|M_{bc}-5.2792|<0.007$~GeV/$c^{2}$ ($2.7\sigma$) for both the electron mode and the muon mode and $-0.06<|\Delta E|<0.04$~GeV for the electron mode and $|\Delta E|<0.040$~GeV for the muon mode. We require ${\cal{LR}}_{cont} > 0.6$ and ${\cal{LR}}_{B\overline{B}} > 0.3$ for all modes except for $K^{0}_{S}\pi^{+}$ and $K^{+}\pi^{0}$ final states, where the selection criterion is tightened to be ${\cal{LR}}_{B\overline{B}} > 0.35$.
The detection efficiencies estimated using MC samples are summarized in Table \ref{tab:results}.
Figure~\ref{fig:dE-Mb} shows the $\Delta E$ vs. $M_{bc}$ scatterplot for each mode.

To determine the signal yields, we must take into account backgrounds from both misidentified leptons and real leptons. The $M_{bc}$ distribution is fitted with the sum of a Gaussian function to represent the signal and two ARGUS functions\cite{argus} plus a Gaussian function to represent the background.
The mean and the width of the Gaussian function are calibrated using $J/\psi K^{+}$ and $J/\psi K^{*0}$ events.
The shape of the background function is determined using data to model the background from misidentified leptons and a 200~fb${}^{-1}$ MC dilepton sample for the real lepton background.
One ARGUS function and one Gaussian function in the background function is introduced to account for the misidentified lepton background.
The other ARGUS function represents the real lepton background and its normalization is a free parameter in the fit.

\begin{figure}
\hspace{6mm}(a) $B^0\to K^{0} e^{+} e^{-}$\hspace{4.36cm}(b) $B^+\to K^{+} e^{+} e^{-}$\\
\vspace*{-10mm}
\begin{center}
\epsfxsize 2.8 truein \epsfbox{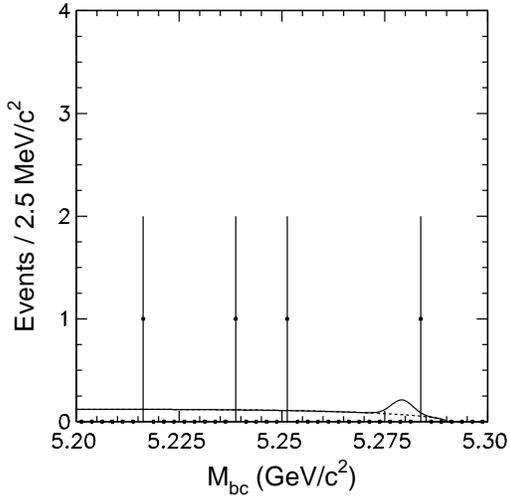}
\hspace{3mm}
\epsfxsize 2.8 truein \epsfbox{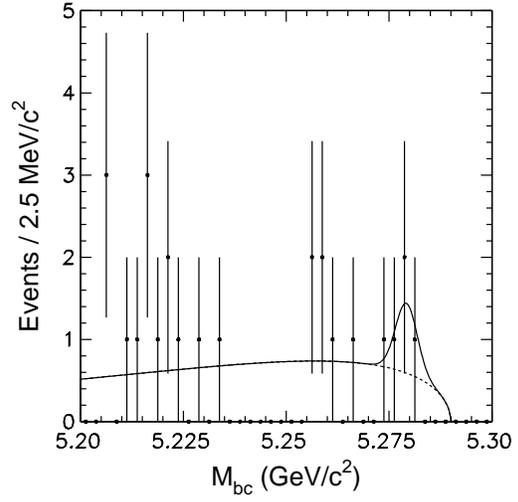}\\
\end{center}
\hspace{6mm}(c)$B^0\to K^{*0} e^{+} e^{-}$\hspace{4.4cm}(d)$B^+\to K^{*+} e^{+} e^{-}$\\
\vspace*{-10mm}
\begin{center}
\epsfxsize 2.8 truein \epsfbox{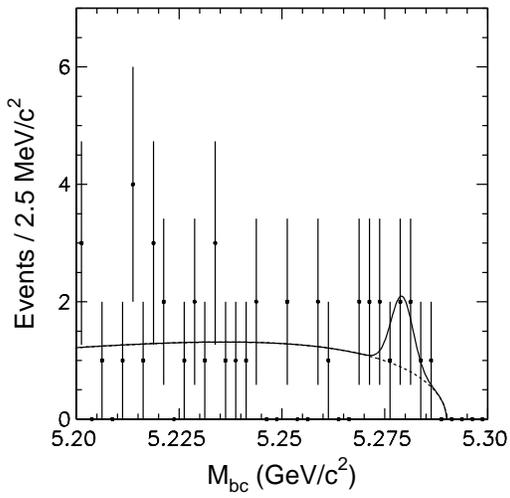}
\hspace{3mm}
\epsfxsize 2.8 truein \epsfbox{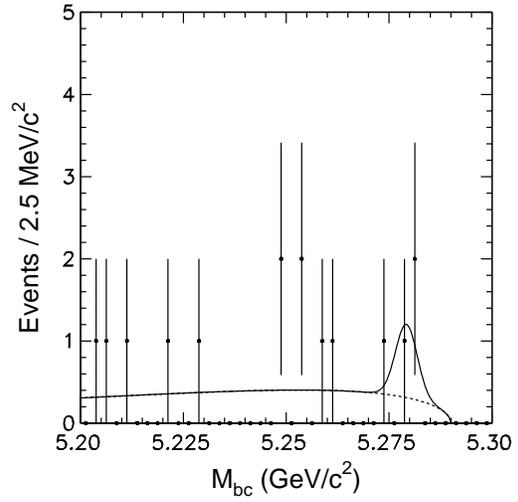}\\
\end{center}
\caption{$M_{bc}$ distributions with fits for (a) $B^0\to K^{0} e^{+} e^{-}$, (b) $B^+\to K^{+} e^{+} e^{-}$, (c) $B^0\to K^{*0} e^{+} e^{-}$ and (d) $B^+\to K^{*+} e^{+} e^{-}$ samples in data.
}
\label{fig:mbproj-ee}
\end{figure}

\begin{figure}
\hspace{6mm}(a) $B^0\to K^{0} \mu^{+} \mu^{-}$\hspace{4.31cm}(b) $B^+\to K^{+} \mu^{+} \mu^{-}$\\
\vspace*{-10mm}
\begin{center}
\epsfxsize 2.8 truein \epsfbox{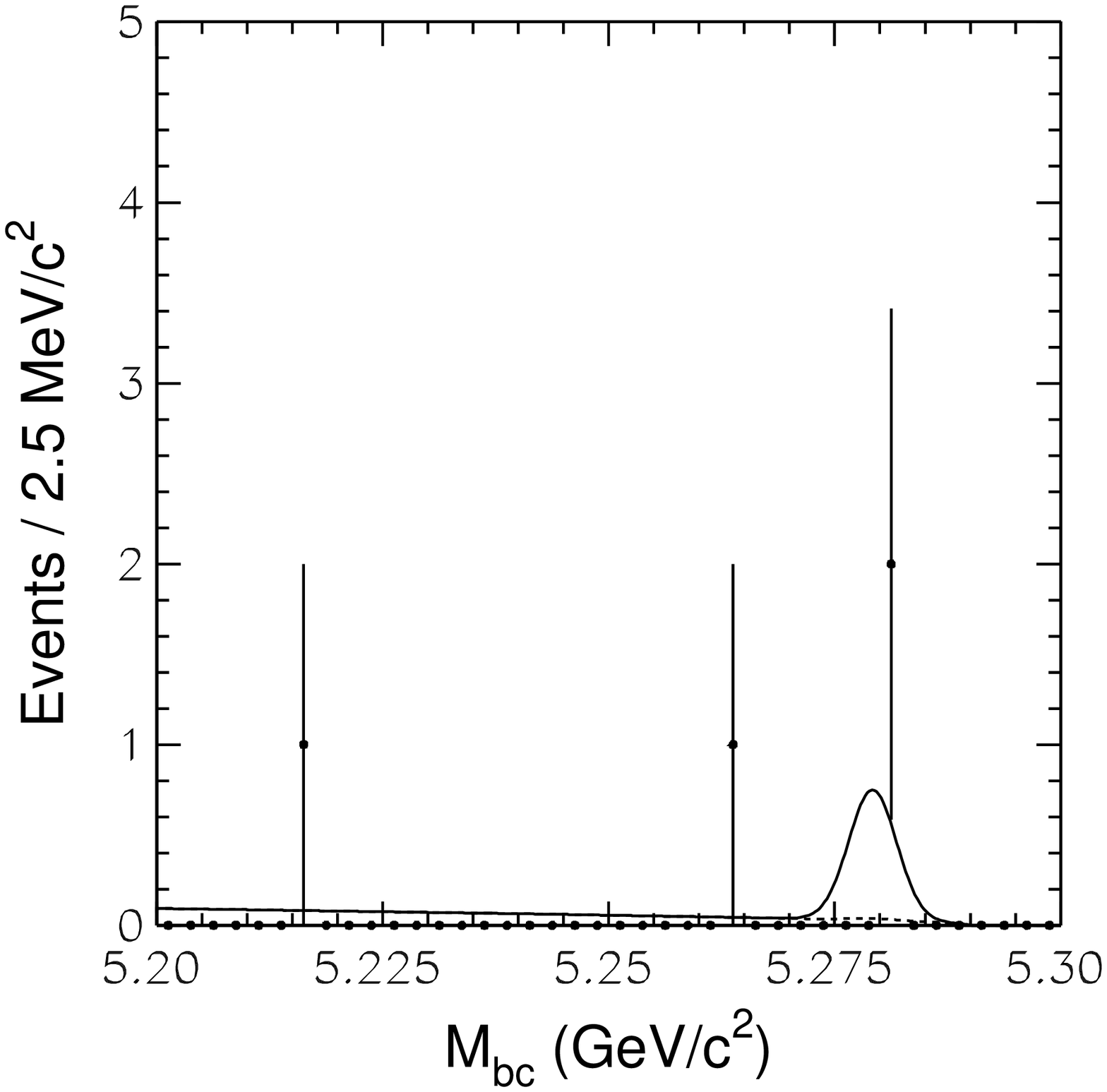}
\hspace{3mm}
\epsfxsize 2.8 truein \epsfbox{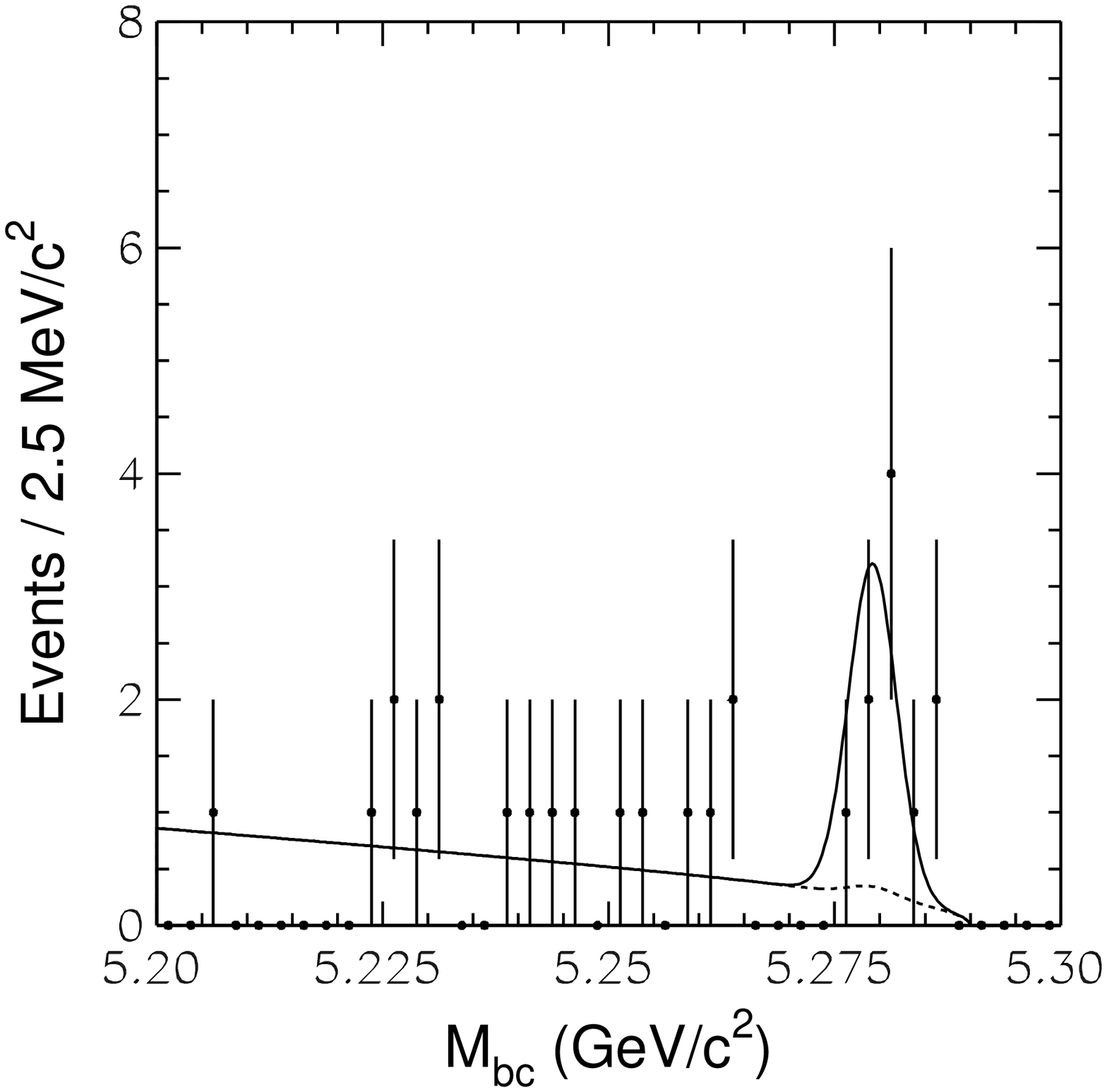}\\
\end{center}
\hspace{6mm}(c)$B^0\to K^{*0} \mu^{+} \mu^{-}$\hspace{4.3cm}(d)$B^+\to K^{*+} \mu^{+} \mu^{-}$\\
\vspace*{-10mm}
\begin{center}
\epsfxsize 2.8 truein \epsfbox{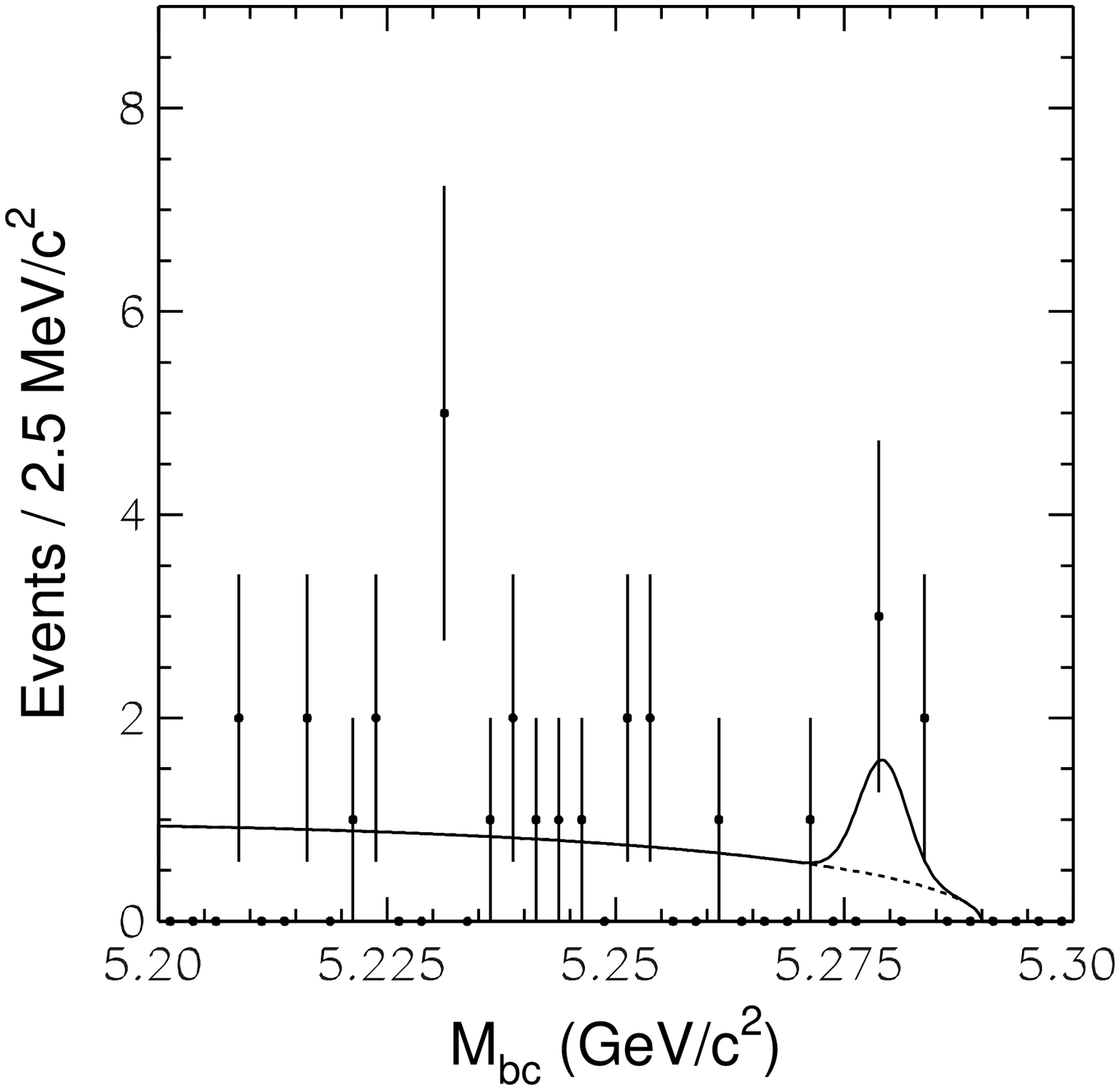}
\hspace{3mm}
\epsfxsize 2.8 truein \epsfbox{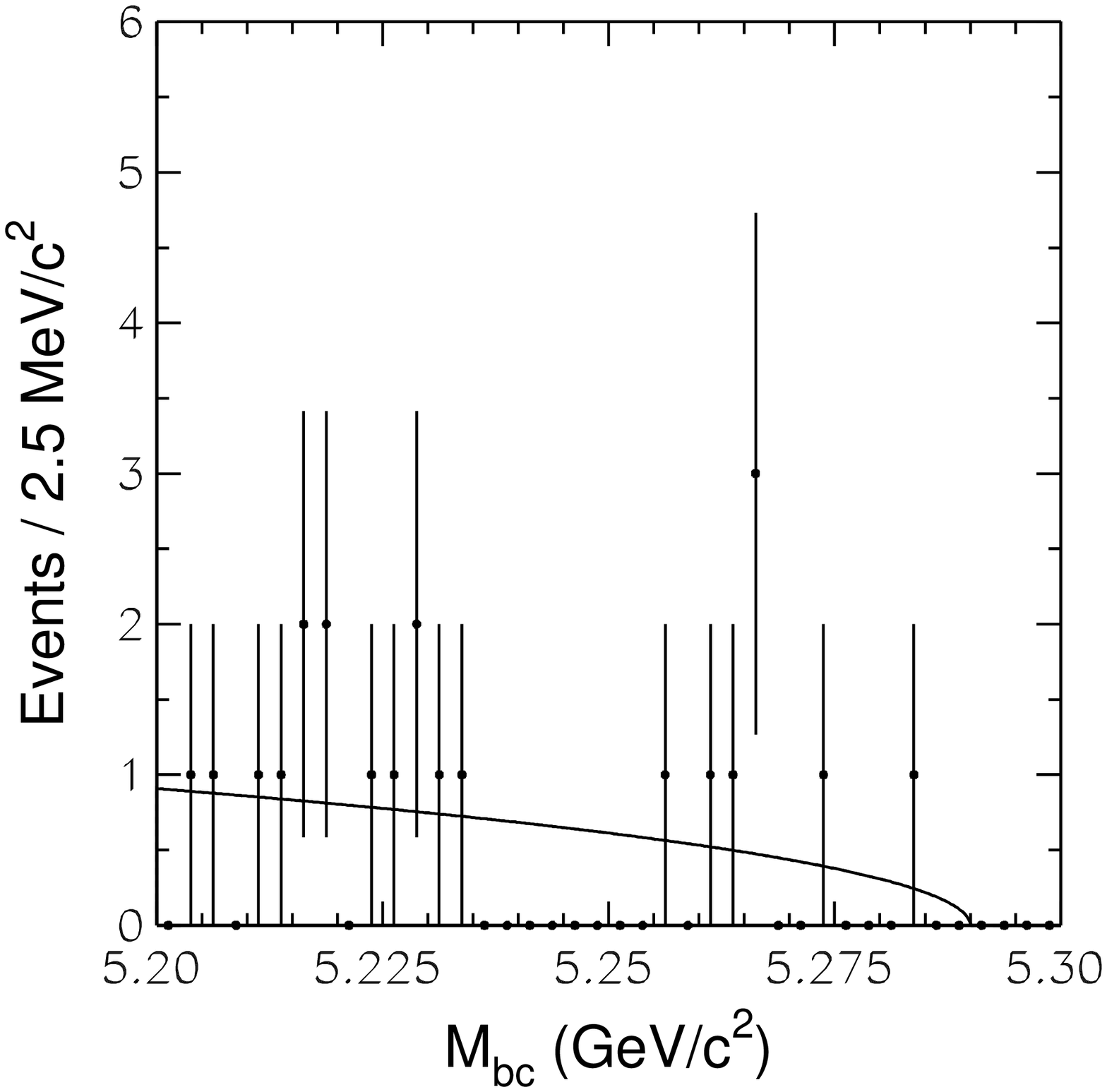}\\
\end{center}
\caption{$M_{bc}$ distributions with fits for (a) $B^0\to K^{0} \mu^{+} \mu^{-}$, (b) $B^+\to K^{+} \mu^{+} \mu^{-}$, (c) $B^0\to K^{*0} \mu^{+} \mu^{-}$ and (d) $B^+\to K^{*+} \mu^{+} \mu^{-}$ samples in data.
}
\label{fig:mbproj-mumu}
\end{figure}

Figures~\ref{fig:mbproj-ee} and \ref{fig:mbproj-mumu} show the $M_{bc}$ distributions after the selection on $\Delta E$.
The fit results for the $e^+e^-$ and $\mu^+\mu^-$ samples are also shown.
The fit results are summarized in Table \ref{tab:results}.

\begin{figure}
\hspace{1mm}(a)\hspace{8cm}(b)\\
\vspace*{-12mm}
\begin{center}
\epsfxsize 3.0 truein \epsfbox{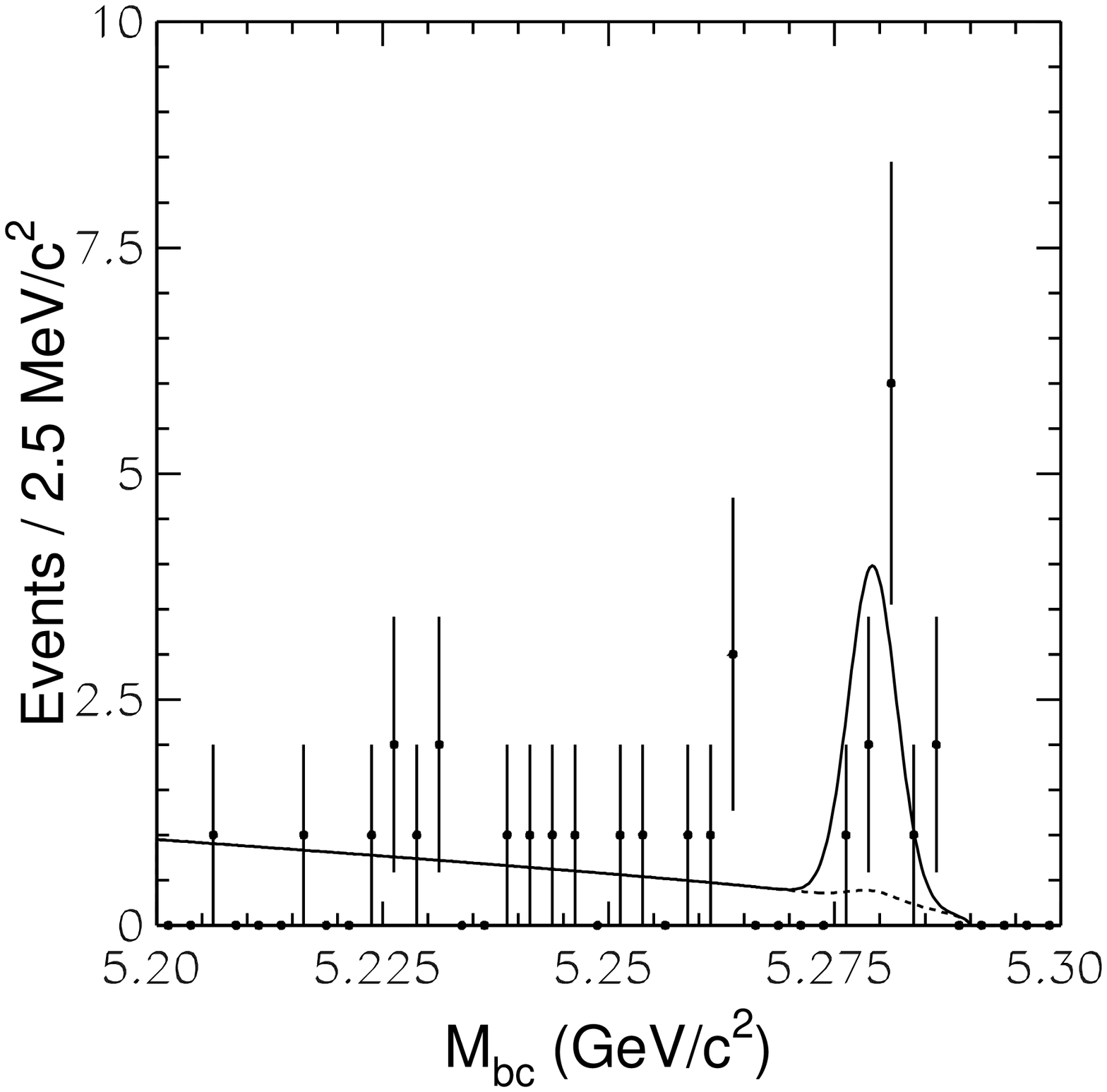}
\hspace{6mm}
\epsfxsize 3.0 truein \epsfbox{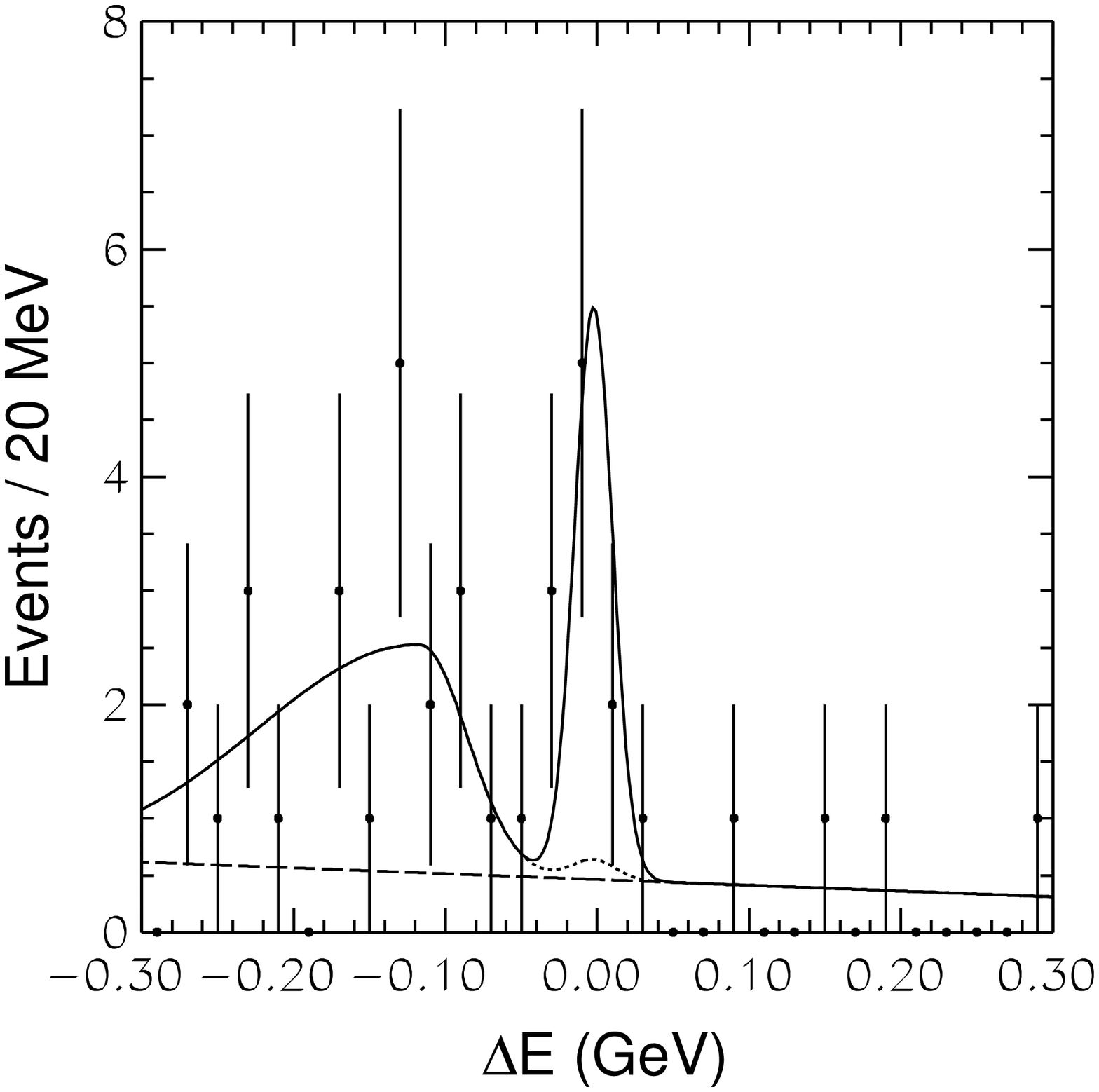}
\end{center}
\caption{ (a) $M_{bc}$ distribution with fits and (b) $\Delta E$ distribution for $B^0\to K^{0} \mu^{+} \mu^{-}$ and $B^+\to K^{+} \mu^{+} \mu^{-}$ candidates combined.
}
\label{fig:kmumu-fit}
\end{figure}

We observe an excess in $B^0\to K^{0} \mu^{+} \mu^{-}$ with a significance of $2.6\sigma$ and in $B^+\to K^{+} \mu^{+} \mu^{-}$ with a significance of $4.1\sigma$. Statistical significance is calculated as $\sqrt{-2\ln{\cal L}_0/{\cal L}_{\rm{max}}}$ where ${\cal L}_{\rm{max}}$ is the maximum likelihood in the $M_{bc}$ fit and ${\cal L}_0$ is the likelihood when the signal yield is constrained to be zero.
When the $M_{bc}$ distribution of $B^0\to K^{0} \mu^{+} \mu^{-}$ and $B^+\to K^{+} \mu^{+} \mu^{-}$ modes are combined, the fit yields $9.53^{+3.74}_{-3.06}$ signal events as shown in Figure~\ref{fig:kmumu-fit} (a).
The statistical significance now increases to 4.8$\sigma$.
We observe a clear peak in the signal region.
Figure~\ref{fig:kmumu-fit} (b) shows the $\Delta E$ distribution of the $B\to K\mu^{+} \mu^{-}$ candidates. The $\Delta E$ fit yield $7.80^{+3.50}_{-2.84}$ is consistent with the yield from the $M_{bc}$ fit.

\begin{figure}
\hspace{0.5mm}(a)\hspace{7.77cm}(b)\\
\vspace*{-13mm}
\begin{center}
\epsfxsize 2.8 truein \epsfbox{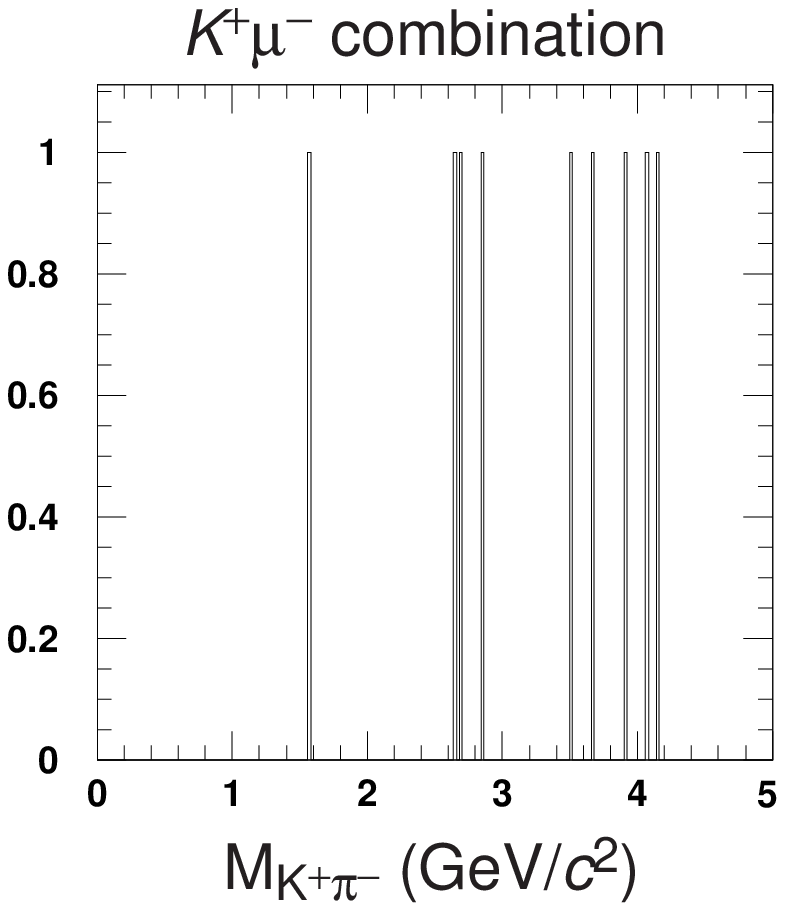}
\hspace{9mm}
\epsfxsize 2.8 truein \epsfbox{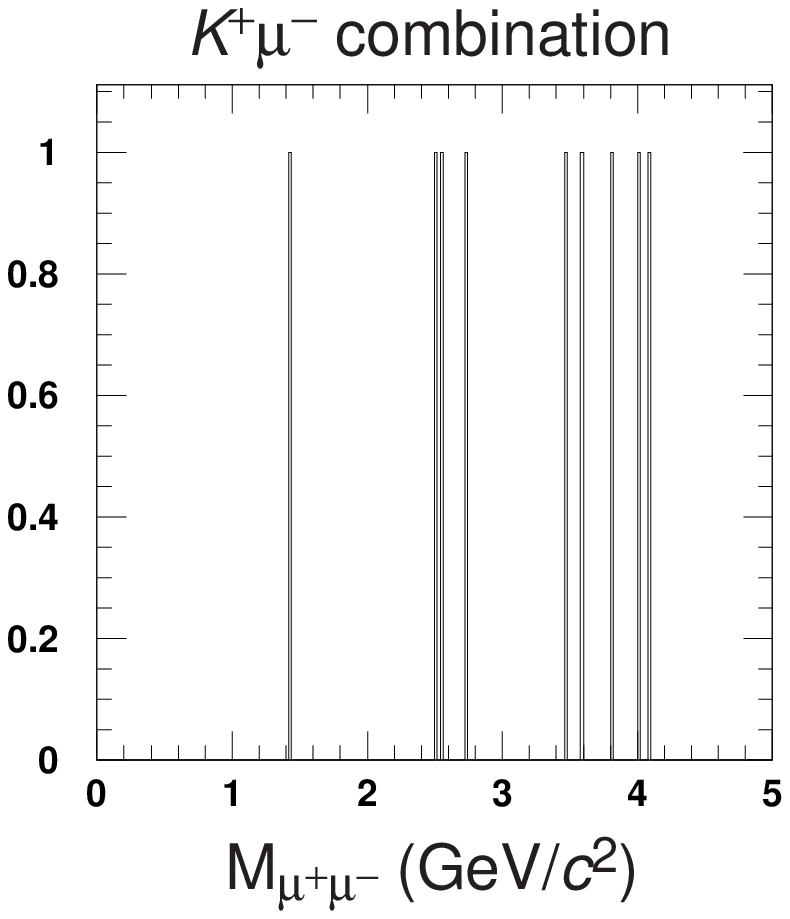}\\
\end{center}
\hspace{1mm}(c)\\
\vspace*{-8mm}
\begin{center}
\epsfxsize 6 truein \epsfbox{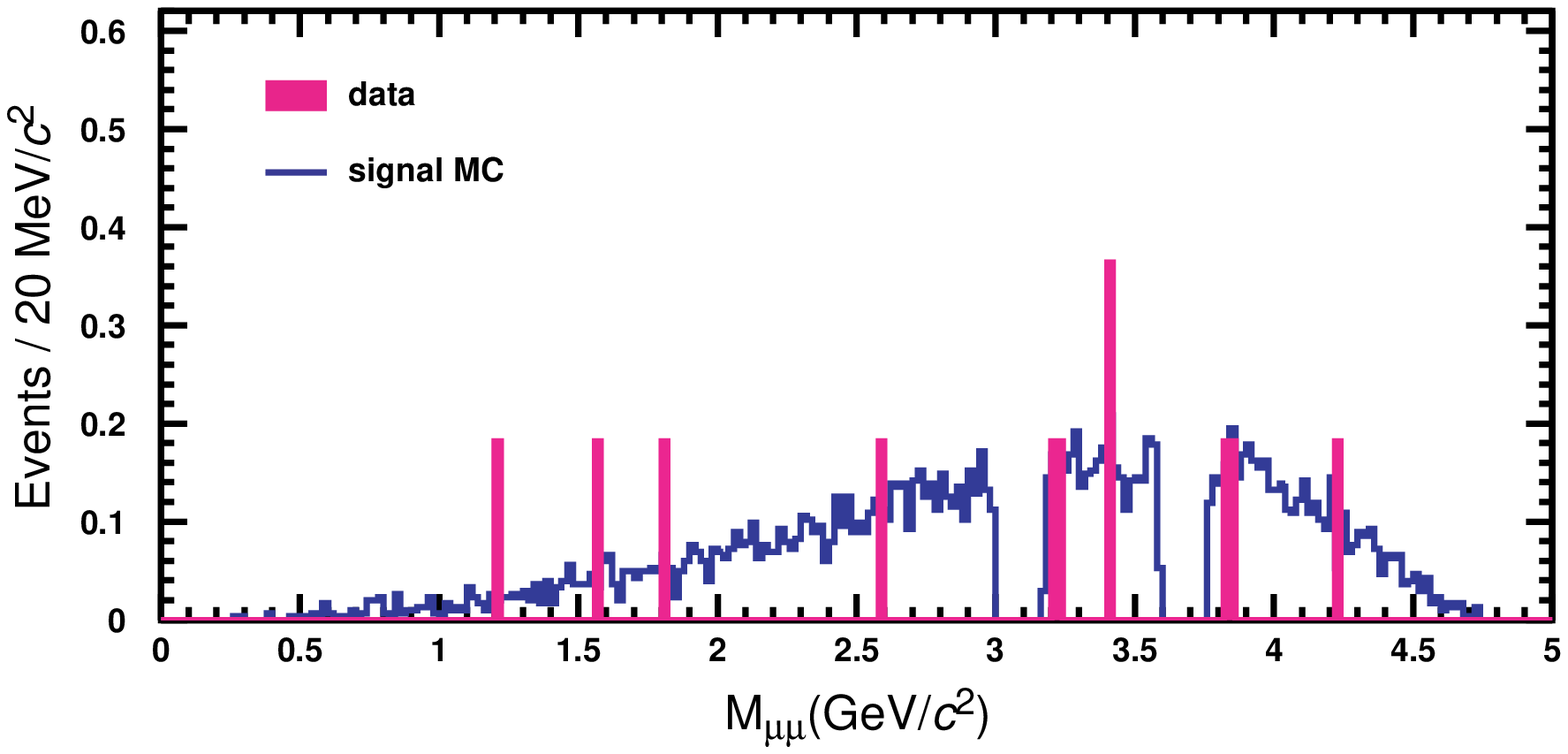}
\end{center}
\caption{
Mass distribution of $K^+\mu^-$ combinations with (a) $K^+\pi^-$ and (b) $\mu^{+} \mu^{-}$ hypotheses.
(c) Dimuon mass distribution of $B\to K \mu^{+} \mu^{-}$ candidates.
}
\label{fig:kmumu-mass}
\end{figure}

The kinematical properties of the $B\to K\mu^{+} \mu^{-}$ events are further examined to check for potential backgrounds.
$B\to K h^+h^-$ decays ($h^{\pm}$ refers hadrons) can contribute to the peak in the $M_{bc}$ distribution when both of hadrons are misidentified as muons.
The $B^{+}\to \overline{D^{0}}\pi^+,~\overline{D^{0}}\to K^{+}\pi^-$ decay chain is expected to be the largest source of this background.
We expect 0.2 events from this source using a 300~fb${}^{-1}$ MC sample.
The $B\to K h^+h^-$ background is also evaluated using data. All  $K h^+h^-$ combinations are weighted by the measured misidentification probability.
This study yields 0.3 $B\to K h^+h^-$ events in the peak region, which is consistent with the MC result.
Another possible background source is double-misidentification of the $B\to J/\psi K,~J/\psi\to \mu^+\mu^-$ decay chain where the kaon and the muon are misidentified as a muon and a kaon, respectively.
Figs.~\ref{fig:kmumu-mass} (a) and (b) show mass distribution of the $K^+\mu^-$ combinations with the $K^+\pi^-$ and the $\mu^+\mu^-$ hypotheses, respectively.
We observe no events in the $D^0$ mass or $J/\psi$ mass region, which confirms the MC expectation.

The $B\to J/\psi X,~J/\psi\to \mu^+\mu^-$ decay chain can be another background source when muon pairs from $J/\psi\to \mu^+\mu^-$ decays avoid the $\psi^{(')}$ veto.
We expect 0.1 events from this background using a $B\to J/\psi X$ MC sample corresponding to 220~fb${}^{-1}$.
Fig.~\ref{fig:kmumu-mass} (c) shows the dimuon mass distribution for the $B\to K\mu^+\mu^-$ candidates.
The hatched histogram shows the data distribution while the solid histogram shows the MC signal distribution.
The data distribution is consistent with the MC expectation.
We observe no events close to the $J/\psi$ or $\psi'$ veto region, which confirms the MC expectation for the background contribution.
To summarize, we observe no indication of a background that could peak in the $M_{bc}$ distribution in the $B\to K \mu^{+} \mu^{-}$ sample.

\subsection{Inclusive Analysis}
The inclusive $B\to X_s\ell^+\ell^-$ decay is reconstructed by combining the $X_s$ system with two oppositely charged leptons.
The $X_s$ system is formed by combining a charged kaon or $K^0_S$ with 0--4 pions in which at most one pion can be neutral.

We reject backgrounds from $B\to\psi^{(')}X_s$ decays by vetoing $-0.6<M_{e^+e^-}-M_{\psi}<0.15$~GeV/$c^2$, $-0.30<M_{e^+e^-}-M_{\psi'}<0.15$~GeV/$c^2$, $-0.35<M_{\mu^+\mu^-}-M_{\psi}<0.2$~GeV/$c^2$ and $-0.30<M_{\mu^+\mu^-}-M_{\psi'}<0.15$~GeV/$c^2$.
These criteria are rather tight since these backgrounds can satisfy the $\Delta E$ requirement described below by adding additional pions and populate the $M_{bc}$ sideband.
We also reject electrons from photon conversions or $\pi^0$ Dalitz decays by requiring $M_{e^+e^-}>0.2$~GeV/$c^2$.

Continuum backgrounds are reduced by requiring $R_2$ to be less than 0.35.
The remaining continuum backgrounds are further suppressed by the requirement $|\cos\theta_{\mathrm{thrust}}|<0.85$ where $\theta_{\mathrm{thrust}}$ is the angle between the thrust axis of the candidate and the thrust axis of the rest of the event.
Continuum events peak at $\pm 1$ while the signal events are uniformly distributed.
After these requirements, the continuum background becomes negligibly small; the requirement of two leptons eliminates almost all $u\overline{u},~d\overline{d},~s\overline{s}$ events.
Most of the remaining background is from $B\overline{B}$ events with two leptons from semileptonic decays of either $B$ or $D$ mesons.

We calculate a likelihood ratio, ${\cal LR}_{\mathrm{K}}$, using the missing energy in the event and the invariant mass of the $X_s$ system.
In the variable $\cos\theta_{K\ell^+}+\cos\theta_{K\ell^-}$, where $\theta_{K\ell}$ is the angle between the charged or neutral kaon (not $X_s$) and the lepton, the background peaks around zero while the signal tends to be negative.
A cut on this variable does not bias the $M_{bc}$ distribution.
This variable is combined with the $B$ flight direction to form another likelihood ratio, ${\cal LR}_{\theta}$.
We require ${\cal LR}_{\mathrm{K}}>0.525$ for the $e^+e^-$ sample and ${\cal LR}_{\mathrm{K}}>0.60$ for the $\mu^+\mu^-$ sample, and ${\cal LR}_{\theta}>0.20$ for both samples, which minimize the expected upper limits.

When we have two or more $B\to X_s\ell^+\ell^-$ candidates in one event, we choose the candidate with minimum $(\Delta E/\sigma_{\Delta E})^2$.
Finally, $|\Delta E|$ is required to be less than 30~MeV.

\begin{figure}
\hspace{9mm}(a) $B\to X_s e^+e^-$\hspace{4.6cm}(b) $B\to X_s \mu^+\mu^-$\\
\vspace*{-5mm}
\begin{center}
\epsfxsize 2.8 truein \epsfbox{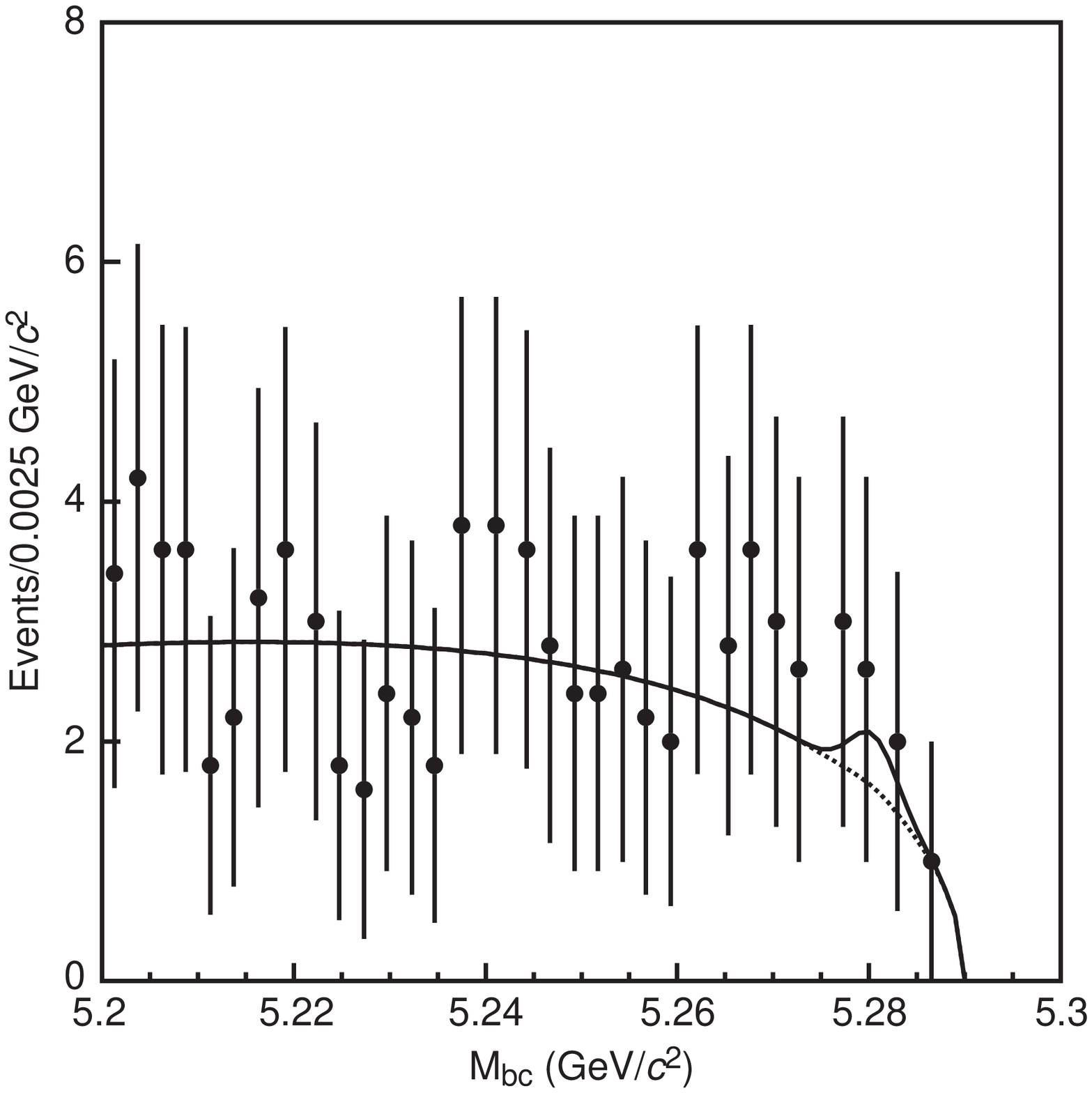}
\hspace{3mm}
\epsfxsize 2.8 truein \epsfbox{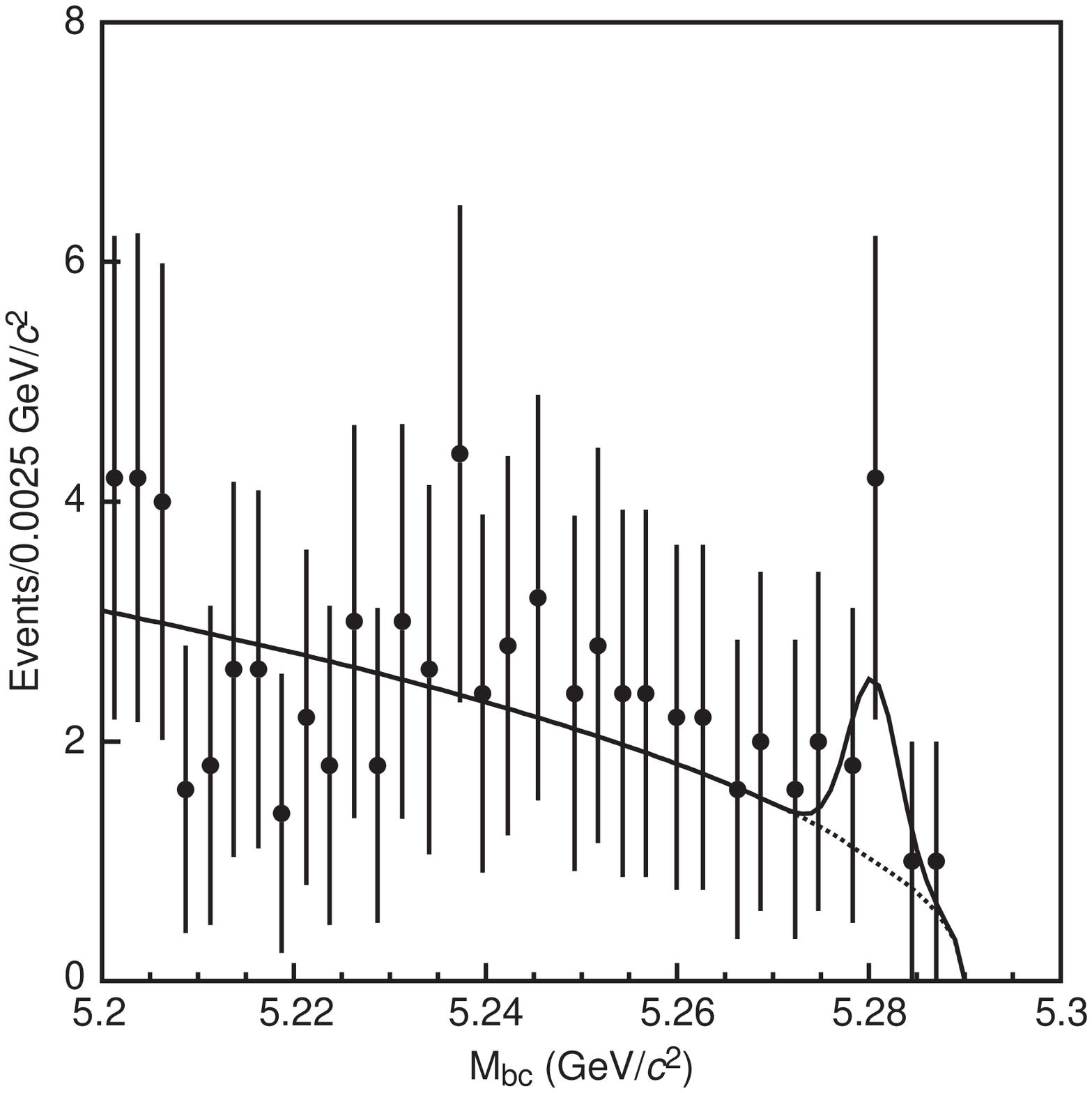}\\
\end{center}
\caption{$M_{bc}$ distributions with fits for (a) $B\to X_s e^+e^-$ sample and (b) $B\to X_s \mu^+\mu^-$ sample in data.
Points with error bar show the data.
The solid curve shows the sum of signal and background while the dotted curve shows the background.}
\label{fig:mbc-inclusive}
\end{figure}

We obtain the signal yield by a fit to the $M_{bc}$ distribution with the sum of a Gaussian function for the signal and an ARGUS function for the background.
The background shape is  determined in this fit while the signal function shape is calibrated using the $B\to J/\psi K$ sample in data.
The fits yield $3.0^{+4.9}_{-4.3}$ for the $B\to X_s e^+e^-$ sample and $11.4^{+5.1}_{-4.8}$ for the $B\to X_s \mu^+\mu^-$ sample.
Figure \ref{fig:mbc-inclusive} shows the $M_{bc}$ distributions with the fit results.
The statistical significance of the $B\to X_s \mu^+\mu^-$ signal is $2.7\sigma$.
Of these 11.4 signal events, 5.9 events are reconstructed as either $B\to K\mu^+\mu^-$ or $B\to K^*\mu^+\mu^-$.

\section{Systematic Uncertainties}
We consider systematic effects from the fit and the efficiency determination.
Uncertainty in the background function is the dominant source of the systematic error.

\subsection{Systematic error due to fit}
The systematic error associated with the signal function is evaluated by varying the mean and the width of the Gaussian function determined from $J/\psi K^{(*)}$ events by $\pm 1 \sigma$.

In the exclusive analysis, the systematic error due to the background function is obtained by varying the ARUGS shape parameter, which was determined from a large MC sample, by $\pm 1 \sigma$.

In the inclusive analysis the $\Delta E$ requirement is varied to estimate the systematic error associated with modeling of the background shape.
We also use the background shape from a MC sample corresponding to 80~fb${}^{-1}$.
The differences from the above results are considered to be systematic errors.

Total systematic errors are calculated by adding all systematic errors in quadrature.
The systematic errors associated with the fit function are shown in the third column of Table \ref{tab:results}.

\subsection{Systematic error due to efficiency}

Systematic uncertainties for the tracking, charged kaon ID, charged pion ID, electron ID, muon ID, $K^{0}_{S}$ detection and $\pi^{0}$ detection efficiencies are 1.5 to 1.7\%, 2.1 to 2.5\%, 0.8\%, 1.8\%, 2.2\%, 8.7\% and 6.8\% per particle, respectively.
For each final state in the inclusive analysis, these systematic uncertainties are included and weighted by the fraction of the final state times efficiency to calculate the average systematic error over all final states.
Table \ref{table:sys} summarizes the systematic errors associated with the efficiency determination.
\begin{table}[htdp]
\caption{Summary of systematic errors in efficiencies.}
(a) Exclusive Analysis.
\begin{center}
\begin{tabular}{lcccccc}
 & \multicolumn{6}{c}{Fractional error (\%)} \\ \cline{2-7}
{Source} & \multicolumn{6}{c}{electron mode / muon mode} \\ \cline{2-7}
                      & $K^{0}$   & $K^{+}$ & $K^{+}\pi^{-}$ & $K^{0}_{S}\pi^{0}$ & $K^{0}_{S}\pi^{+}$ & $K^{+}\pi^{0}$  \\ \hline
Tracking              & 3.4       & 5.1     & 6.4            & 3.4                & 4.9                & 4.9             \\
Kaon ID               & -         & 2.5     & 2.1            & -                  & -                  & 2.1             \\
Pion ID               & -         & -       & 0.8            & -                  & 0.8                & -               \\
Lepton ID             & 3.6/4.4   & 3.6/4.4 & 3.6/4.4        & 3.6/4.4            & 3.6/4.4            & 3.6/4.4         \\
$K^{0}_{S}$ detection & 8.7       & -       & -              & 8.7                & 8.7                & -               \\
$\pi^{0}$ detection   & -         & -       & -              & 6.8                & -                  & 6.8             \\
BG suppression        & 2.3       & 2.3     & 2.3            & 2.3                & 2.3                & 2.3             \\
MC statistics         & 1.9/1.6   & 1.5/1.3 & 2.2/1.6        & 4.0/3.7            & 2.5/2.3            & 3.5/2.9         \\ \hline
total                 & 10.4/10.7 & 7.3/7.7 & 8.3/8.6        & 13.0/13.1          & 11.2/11.4          & 10.3/10.4       \\
\end{tabular}
\end{center}
(b) Inclusive Analysis.
\begin{center}
\begin{tabular}{lcc}
\raisebox{-1ex}[0pt]{Source} & \multicolumn{2}{c}{Fractional error (\%)} \\ \cline{2-3}
					& $B\to X_s e^+e^-$ & $B\to X_s \mu^+\mu^-$ \\ \hline
Tracking			& $\pm8.1$		& $\pm7.7$	\\
Kaon ID				& $\pm0.8$		& $\pm0.9$	\\
Lepton ID			& $\pm3.6$		& $\pm4.4$	\\
$\pi^0$ detection	& $\pm1.5$		& $\pm0.9$	\\
MC statistics		& $\pm8.1$      & $\pm10$   \\
Model				& ${}^{+22}_{-26}$ & ${}^{+24}_{-22}$ \\
\hline
Total				& ${}^{+25}_{-29}$ & ${}^{+27}_{-25}$ \\ 
\end{tabular}
\end{center}
\label{table:sys}
\end{table}%

The uncertainty in the reconstruction efficiency due to signal modeling in the inclusive analysis is estimated by varying the branching fractions by the errors shown in Tables \ref{tab:brpred} and \ref{tab:brpredinc}.
Model dependence of the efficiency is also considered by using branching fractions from different models.
Since exclusive modes have much higher efficiencies, uncertainties in their branching fractions are the dominant source of the systematic error.

\section{Results}

For the modes with more than $2.5\sigma$ significance, we quote branching fractions. We find,
\begin{center}
	${\cal B}(B^0 \to K^0 \mu^{+} \mu^{-}) = (0.93^{+0.88}_{-0.55}{}\pm{0.10}) \times 10^{-6}$,

	${\cal B}(B^+ \to K^+ \mu^{+} \mu^{-}) = (1.01^{+0.46}_{-0.36}{}^{+0.14}_{-0.16}) \times 10^{-6}$,
\end{center}
where the first and second errors are statistical and systematic, respectively.
We combine neutral and charged $B$-meson results for $B\to K \mu^{+} \mu^{-}$ modes and obtain the combined branching fraction,
\begin{center}
	${\cal B}(B \to K \mu^{+} \mu^{-}) = (0.99^{+0.39}_{-0.32}{}^{+0.13}_{-0.15}) \times 10^{-6}$.
\end{center}

In order to calculate upper limits for signal yields in the presence of background, we employ the unified approach
of Feldman and Cousins.
The upper limits for the branching fractions are calculated using the upper limit of the signal yield and lower limit of the efficiency to obtain conservative upper limits.
We obtain upper limits at 90\% confidence level, which are given in Table \ref{tab:results}.

\begin{table}[htpb]
\caption{Summary of the fit results and branching fractions. 
Number of events observed in the signal box, number of signal and background events estimated from the $M_{bc}$ fit. 
The first error in the signal yield and branching fraction is statistical and the second one is systematic.}

\vspace*{5mm}
(a) Exclusive Analysis.
\begin{center}
\begin{tabular}{lccccccc} 
\raisebox{-1ex}[0pt]{mode} & observed & signal yield & background & efficiency & ${\cal{B}}$ & U.L.  &  signif. \\
& events  & & & [\%] & $[\times10^{6}]$ &  $[\times10^{6}]$  & \\ \hline
$K^{0} e^{+} e^{-}      $ & 1        & $0.38^{+1.39}_{-0.38}{}^{+0.56}_{-0.38}$ & 0.62       & $5.51 \pm 0.58$ & -                                             & 2.78               & -          \\ 
$K^{+} e^{+} e^{-}      $ & 5        & $2.25^{+2.49}_{-1.82}{}^{+1.03}_{-1.36}$ & 2.75       & $21.6 \pm 1.6 $ & -                                             & 1.32               & -          \\ 
$K e^{+} e^{-}          $ & 6        & $2.63^{+2.70}_{-2.02}{}^{+1.16}_{-1.42}$ & 3.37       & $27.1 \pm 2.2 $ & -                                             & 1.19               & -          \\ \hline
$K^{*0} e^{+} e^{-}$      & 9        & $3.22^{+3.07}_{-2.36}{}^{+1.12}_{-1.24}$ & 5.78       & $6.58 \pm 0.57$ & -                                             & 5.63               & -          \\ 
$K^{*+} e^{+} e^{-}$      & 4        & $2.36^{+2.27}_{-1.60}{}^{+0.40}_{-0.48}$ & 1.64       & $3.06 \pm 0.32$ & -                                             & 8.57               & -          \\ 
$K^{*} e^{+} e^{-}$       & 13       & $5.52^{+3.68}_{-2.98}{}^{+1.15}_{-1.26}$ & 7.48       & $9.64 \pm 0.89$ & $2.73^{+1.82}_{-1.47}{}^{+0.62}_{-0.67}$      & 5.07               & 2.10          \\ \hline\hline
$K^{0} \mu^{+} \mu^{-}  $ & 2        & $1.88^{+1.77}_{-1.11}{}^{+0.04}_{-0.06}$ & 0.12       & $6.45 \pm 0.69$ & $0.93^{+0.88}_{-0.55}{}\pm{0.10}$      & 3.21               & 2.59          \\
$K^{+} \mu^{+} \mu^{-}  $ & 9        & $7.57^{+3.42}_{-2.74}{}^{+0.90}_{-1.02}$ & 1.43       & $23.9 \pm 1.8 $ & $1.01^{+0.46}_{-0.36}{}^{+0.14}_{-0.16}$      & -                  & 4.09       \\ 
$K \mu^{+} \mu^{-}      $ & 11       & $9.53^{+3.74}_{-3.06}{}^{+0.93}_{-1.16}$ & 1.47       & $30.4 \pm 2.5 $ & $0.99^{+0.39}_{-0.32}{}^{+0.13}_{-0.15}$               & -                  & 4.76       \\ \hline
$K^{*0} \mu^{+} \mu^{-} $ & 6        & $3.03^{+2.58}_{-1.89}{}^{+0.83}_{-1.05}$ & 2.97       & $8.33 \pm 0.72$ & -                                             & 3.90               & -          \\
$K^{*+} \mu^{+} \mu^{-} $ & 2        & $0.00^{+0.86}_{-0.00}{}^{+0.00}_{-0.00}$ & 2.00       & $3.47 \pm 0.38$ & -                                             & 6.10               & -          \\ 
$K^{*} \mu^{+} \mu^{-} $  & 8        & $2.84^{+2.85}_{-2.14}{}^{+1.25}_{-1.46}$ & 5.16       & $11.8 \pm 1.1$  & -                                             & 3.01               & -          \\ 
\end{tabular}
\end{center}
(b) Inclusive Analysis.
\begin{center}
\begin{tabular}{lccccccc} 
\raisebox{-1ex}[0pt]{mode} & observed & signal yield & background & efficiency & ${\cal{B}}$ & U.L.  &  signif. \\
& events  & & & [\%] & $[\times10^{6}]$ &  $[\times10^{6}]$  & \\\hline
$X_s e^{+} e^{-}$ 		& 33 & $3.0^{+4.9}_{-4.3}{}^{+0.56}_{-0.38}$ & 30 & $3.68 ^{+1.05}_{-0.93}$ & -  & 10.2 & -  \\
$X_s \mu^{+} \mu^{-}$	& 26 & $11.4^{+5.1}_{-4.8}{}^{+1.03}_{-1.36}$ & 14.6 & $2.66^{+0.68}_{-0.73}$  & $6.8^{+3.1}_{-2.9}{}^{+2.0}_{-2.4}$ & 19.9  & 2.7 \\ 
\end{tabular} 
\end{center}
\label{tab:results}
\end{table}

\section{Conclusions}
We have observed evidence for the electroweak penguin decay $B\to K\mu^+\mu^-$.
The preliminary branching fraction for this decay mode is,
\begin{center}
	${\cal B}(B \to K \mu^{+} \mu^{-}) = (0.99^{+0.39}_{-0.32}{}^{+0.13}_{-0.15}) \times 10^{-6}$.
\end{center}
This value is consistent with theoretical predictions \cite{Ali} \cite{Greub} \cite{Melikhov}. 
The errors are dominated by statistics, which can be improved in the coming runs. We report 90\% confidence level upper limits of branching fractions for the following exclusive and inclusive $B\to X_s \ell^+\ell^-$ decays. 
\begin{center}
	${\cal B}(B \to K e^{+} e^{-}) < 1.2 \times 10^{-6}$,\\
	${\cal B}(B \to K^* e^{+} e^{-}) < 5.1 \times 10^{-6}$,\\
	${\cal B}(B \to K^* \mu^{+} \mu^{-}) < 3.0 \times 10^{-6}$,\\
	${\cal B}(B \to X_{s} e^{+} e^{-}) < 10.1 \times 10^{-6}$,\\
	${\cal B}(B \to X_{s} \mu^{+} \mu^{-}) < 19.1 \times 10^{-6}$.
\end{center}
These values are close to the SM predictions. 
The results reported here are preliminary.


\section{Acknowledgements}

We wish to thank the KEKB accelerator group for the excellent
operation of the KEKB accelerator. We acknowledge support from the
Ministry of Education, Culture, Sports, Science, and Technology of Japan
and the Japan Society for the Promotion of Science; the Australian
Research
Council and the Australian Department of Industry, Science and
Resources; the Department of Science and Technology of India; the BK21
program of the Ministry of Education of Korea and the CHEP SRC
program of the Korea Science and Engineering Foundation; the Polish
State Committee for Scientific Research under contract No.2P03B 17017;
the Ministry of Science and Technology of Russian Federation; the
National Science Council and the Ministry of Education of Taiwan; the
Japan-Taiwan Cooperative Program of the Interchange Association; and
the U.S. Department of Energy.
A.I. would like to acknowledge support from Research Fellowships of the Japan Society 
for the Promotion of Science for Young Scientists.

\end{document}